\documentclass{new_tlp}
\usepackage{mathptmx}
\usepackage{url}
\usepackage{amsmath}
\usepackage{amsfonts}
\usepackage{xcolor}
\usepackage{graphicx}
\usepackage{epsfig} 
\usepackage{multirow}
\usepackage{xspace}
\usepackage{todonotes}  
\usepackage{color}
\usepackage{asplisting}
\usepackage{listings}
\usepackage{comment}
\usepackage{enumerate}
\usepackage{booktabs}

\usepackage{tikz}
\usepackage{pgfplots}
\pgfplotsset{compat=1.16}

\usepackage{appendix}

\pgfplotstableread{
1	3	3	7	3
2	5	3	7	3
3	6	2	8	3
4	7	3	12	3
5	4	2	5	3
6	5	2	3	3
7	4	3	4	3
8	5	3	4	3
9	8	3	1	3
10	6	2	1	3
11	3	3	3	3
12	2	3	2	3
13	2	2	3	3
14	3	3	5	3
15	2	3	2	3
16	1	3	1	3
17	1	3	3	3
18	2	3	0	3
19	1	3	2	3
20	0	3	0	3
21	0	3	1	3
22	0	3	3	3
23	0	2	0	3
24	1	3	0	3
25	0	3	0	3
26	0	2	1	3
}\dataset

\pgfplotstableread{
1	4	3	1	2	1
2	2	1	2	1	2
3	1	1	3	1	1
4	1	4	2	2	2
5	1	1	1	1	2
6	3	1	3	2	2
7	1	0	3	1	2
8	2	1	1	2	1
9	1	1	1	2	2
10	2	1	3	1	2
11	5	3	1	2	2
12	2	1	1	2	1
13	4	4	2	2	2
14	3	1	1	1	2
15	3	1	3	2	2
16	2	1	2	2	2
17	1	3	3	2	2
18	1	4	1	2	2
19	2	3	3	2	2
20	3	3	1	2	2
21	1	1	3	2	2
22	1	1	3	2	2
23	1	3	1	2	2
24	1	1	3	1	2
25	1	1	1	2	2
26	1	3	1	2	2
}\weeklydataset

\pgfplotstableread{
1	10	1   8   4   
2	5	2   8   2
3	5	1   6   1
4	3   2   9   1
5	3	2   2   1
6	1	2   3   3
7	2	2   3   1
8	2	1   4   2
9	2	2   4   1
10	3	2   4   2
11	0	2   1   5
12	1	1   1   2
13	1	2   1   4
14	1	2   1   3
15	0	2   2   3
16	0	2   0   2
17	0	2   2   1
18	1	2   0   1
19	0	2   1   2
20	0	2   0   3  
21	0	2   0   1 
22	1   2   1   1
23	0	2   1   1
24	0	2   0   1
25	1	2   1   1
26	0	2   0   1
}\weeklydailydataset

\pgfplotstableread{
1	3	3	3
2	3	3	3
3	3	3	3
4	3	3	3
5	3	3	3
6	3	3	2
7	3	3	3
8	3	3	2
9	2	3	2
10	3	3	2
11	3	3	3
12	3	2	3
13	3	2	3
14	3	2	2
15	0	2	3
16	3	2	3
17	1	2	2
18	2	2	2
19	3	3	2
20	1	3	2
21	0	3	3
22	3	1	2
23	3	1	3
24	3	2	2
25	3	3	2
}\extendedctsdataset

\pgfplotstableread{
1     1    4    3    5    4    0    3    8    5
2     3    6    4    5    4    2    6    9    5
3     0    0    0    1    3    5    0    1    0
4     0    0    1    0    5    0    0    1    0
5     1    1    0    0    1    1    0    0    0
6     0    1    1    0    0    0    1    1    0
7     0    0    1    0    1    1    0    0    0
8     0    0    0    0    1    3    0    0    0
9     2    1    0    0    0    0    0    0    0
10    0    2    2    1    1    1    0    1    1
11    1    2    0    0    2    1    0    1    0
12    0    1    1    0    0    0    2    4    0
13    0    0    1    0    0    3    0    1    0
14    0    0    3    0    0    0    1    0    0
15    0    1    1    0    2    1    0    0    0
16    1    1    3    0    1    2    1    0    0
17    0    2    0    0    2    0    0    1    0
18    1    2    2    1    0    3    1    1    0
19    0    0    5    0    0    1    0    1    1
20    0    3    1    1    1    0    0    1    1
21    0    1    0    0    2    2    0    2    1
22    0    0    0    0    2    1    0    1    0
23    1    0    1    0    1    1    0    0    3
24    1    2    0    0    0    0    0    1    0
25    0    0    1    0    0    3    1    0    1
26    1    0    2    1    1    1    0    2    0
27    0    0    0    0    1    1    0    1    1
28    0    1    1    1    0    1    0    1    2
29    1    0    1    0    2    1    0    1    1
30    1    0    0    0    2    2    0    2    2
31    1    1    2    0    1    1    0    0    5
32    0    0    0    0    1    2    0    1    2
33    0    0    0    0    0    1    0    0    10 
34    0    11   9    0    2    5    0    0    5
}\prioritydataset
\pgfplotstableread{
1     1    4    3    5    4    0    3    8    5
2     3    6    4    5    4    2    6    9    5
3     0    0    0    1    3    5    0    1    0
4     0    0    1    0    5    0    0    1    0
5     1    1    0    0    1    1    0    0    0
6     0    1    1    0    0    0    1    1    0
7     0    0    1    0    1    1    0    0    0
8     0    0    0    0    1    3    0    0    0
9     2    1    0    0    0    0    0    0    0
10    0    2    2    1    1    1    0    1    1
11    1    2    0    0    2    1    0    1    0
12    0    1    1    0    0    0    2    4    0
13    0    0    1    0    0    3    0    1    0
14    0    0    3    0    0    0    1    0    0
15    0    1    1    0    2    1    0    0    0
16    1    1    3    0    1    2    1    0    0
17    0    2    0    0    2    0    0    1    0
18    1    2    2    1    0    3    1    1    0
19    0    0    5    0    0    1    0    1    1
20    0    3    1    1    1    0    0    1    1
21    0    1    0    0    2    2    0    2    1
22    0    0    0    0    2    1    0    1    0
23    1    0    1    0    1    1    0    0    3
24    1    2    0    0    0    0    0    1    0
25    0    0    1    0    0    3    1    0    1
26    1    0    2    1    1    1    0    2    0
27    0    0    0    0    1    1    0    1    1
28    0    1    1    1    0    1    0    1    2
29    1    0    1    0    2    1    0    1    1
30    1    0    0    0    2    2    0    2    2
31    1    1    2    0    1    1    0    0    5
32    0    0    0    0    1    2    0    1    2
33    0    0    0    0    0    1    0    0    10 
34    0    11   9    0    2    5    0    0    5
}\prioritytest

\definecolor{darkgreen}{RGB}{0,60,0}
\definecolor{darkgray}{RGB}{80,80,80}

\lstnewenvironment{asp}
{
    \lstset{
		language=asp,
		showstringspaces=false,
		formfeed=\newpage,
		tabsize=4,
		commentstyle=\color{darkgreen},
		basicstyle=\ttfamily\scriptsize,
		keywordstyle=\color{blue},
		numbers=left,
		breaklines=true,
		literate={~} {$\sim$}{1}
	}
}
{
}

\lstnewenvironment{asp22}
{
    \lstset{
		language=asp,
		showstringspaces=false,
		formfeed=\newpage,
		tabsize=4,
		commentstyle=\color{darkgreen},
		basicstyle=\ttfamily\scriptsize,
		keywordstyle=\color{blue},
		numbers=left,
		breaklines=true,
		literate={~} {$\sim$}{1},
		firstnumber=22
	}
}
{
}

\newcommand{\clingo}{\textsc{clingo}\xspace}

\newtheorem{definition}{Definition}

\title[An ASP-based Solution to the Chemotherapy Treatment Scheduling problem]{An ASP-based Solution to the \\ Chemotherapy Treatment Scheduling problem}
\author[Dodaro et al.] {
CARMINE DODARO\\
University of Calabria, Italy\\
\email{dodaro@mat.unical.it}
\and
GIUSEPPE GALAT\`A\\
SurgiQ srl, Italy\\
\email{giuseppe.galata@surgiq.com}
\and
ANDREA GRIONI\\
San Martino Hospital, Italy\\
\email{andrea.grioni@hsanmartino.it}
\and
MARCO MARATEA, MARCO MOCHI\\
University of Genoa, Italy\\
\email{name.surname@unige.it}
\and 
IVAN PORRO\\
SurgiQ srl, Italy\\
\email{ivan.porro@surgiq.com}
}
\jdate{July 2020}
\pubyear{2020}
\pagerange{\pageref{firstpage}--\pageref{lastpage}}
\doi{S1471068401001193}

\begin{document}
\label{firstpage}
\maketitle
\begin{abstract}
The problem of scheduling chemotherapy treatments in oncology clinics is a complex problem, given that the solution has to satisfy (as much as possible) several requirements such as the cyclic nature of chemotherapy treatment plans, maintaining a constant number of patients, and the availability of resources, e.g., treatment time, nurses, and drugs. At the same time, realizing a satisfying schedule is of upmost importance for obtaining the best health outcomes. 

In this paper we first consider a specific instance of the problem which is employed in the San Martino Hospital in Genova, Italy, and present a solution to the problem based on Answer Set Programming (ASP). Then, we enrich the problem and the related ASP encoding considering further features often employed in other hospitals, desirable also in S. Martino, and/or considered in related papers. 
Results of an experimental analysis, conducted on the real data provided by the San Martino Hospital, show that ASP is an effective solving methodology also for this important scheduling problem.
Under consideration for acceptance in TPLP.
\end{abstract}


\section{Introduction}
The Chemotherapy Treatment Scheduling (CTS) \cite{hahn-goldberg_dynamic_2014,huang_alternative_2017,huggins_2014,sevinc_algorithms_2013} problem consists of computing a schedule for patients requiring chemotherapy treatments.
The CTS problem is a complex problem for oncology clinics since it involves multiple resources and aspects, including the availability of nurses, chairs, and drugs.
Chemotherapy treatments have a cyclic nature, where the number and the duration of each cycle depend on the different types of cancer and the stage of the disease.
Moreover, treatments may have different priorities that must be taken into account when computing a solution.
A proper solution to the CTS problem is thus crucial for improving the degree of satisfaction of the main actors on the problem, i.e., patients and nurses, and for a better management of the resources. 
Various studies, also in the context of the COVID19 emergency \cite{kumar_treatment_2020,sud_collateral_2020}, have shown how delays in cancer surgeries and treatments have a significant adverse impact on patient survival. This impact varies depending on the aggressiveness of the cancer, thus stressing the importance of developing a model capable of efficiently prioritize patients. 

Complex combinatorial problems, possibly involving optimizations, such as the CTS problem, are usually the target applications of AI languages such as Answer Set Programming (ASP).
Indeed, ASP has been successfully employed for solving hard combinatorial problems in several research areas,
and it has been also employed to solve many scheduling problems
also in industrial contexts (see, e.g.,  the work by  \citeN{DBLP:journals/aim/ErdemGL16}, \citeN{DBLP:journals/ki/FalknerFSTT18}, \citeN{DBLP:journals/ki/Schuller18}, \citeN{DBLP:conf/aiia/AlvianoBCDGKMMM20} for detailed descriptions of ASP applications).

In this paper, we apply ASP for solving the CTS problem. We first consider a specific instance of the problem which is employed in the San Martino Hospital in Genova, Italy (Section \ref{sec:spec} and \ref{sec:enc}). The problem of the San Martino Hospital consists of assigning a chair or a bed and an hour of treatment to each patient for each fixed day. Some patient requires a bed and the solution has to meet as much request of beds as possible. As an optimization at the moment not implemented in the San Martino Hospital, but highly desired, our solution assigns the patients in such a way to have, as much as possible, the same number of patients during the blood collection performed before the treatment. Moreover, we consider a planning horizon of one week other than the daily actually employed in the hospital. Then, we enrich the problem and the related ASP encoding considering features often employed in other hospitals, considered in related papers \cite{turkanscheduling2010,turkanscheduling2010,Heshmat2021,huggins_2014,DBLP:conf/aiia/DodaroGMP18}, and/or desired by the S. Martino Hospital, i.e., we  explicitly consider the availability of nurses and drugs or a limit to the starting time of treatments (Section \ref{sec:extendedcts}). Both encoding are evaluated on real data of the San Martino Hospital (Section \ref{sec:experiments}): results using the state-of-the-art ASP solver {\sc clingo} \cite{DBLP:journals/ai/GebserKS12} show that ASP is an effective solving methodology for solving all these variants of the presented CTS problem. Focusing on the encoding of the San Martino Hospital, we were able to obtain better schedule w.r.t. the ones that have been implemented in practice, and to obtain very satisfying solutions (i.e., verified by a domain expert) in short time for planning horizons of 1 week instead of one day.

Further, we investigate rescheduling solutions for the weekly real world problem for dealing with situations in which the original schedule cannot be implemented, due to patients that report unavailability (Section \ref{sec:resch}). Goal of the rescheduling is to postpone the canceled registrations, minimizing the changes to the planned schedule while of course still satisfying all the constraints that are in place. Results on some scenario with increasing number of canceled registrations and changed regimen show that rescheduling is done efficiently, in short timings in line with the need of practical applications. 


%

\section{Problem Description}
\label{sec:spec}
In this section, we first present the CTS problem as implemented at the San Martino Hospital in Genova, and then its mathematical formulation.
\subsection{CTS in the real world}
The CTS problem consists of scheduling appointments to a given day for patients requiring chemotherapy treatments, and to assign to each patient a chair or a bed for the whole duration of the session if required.
%
Patients can need different treatments and we identify four phases for each of them:
1) the registration to the Hospital reception;
2) a blood collection;
3) a medical check; and
4) the therapy.
The duration of each phase can be different for each patient.
Moreover, phases 1 and 4 are mandatory, whereas phases 2 and 3 are optional. However, if a patient is involved in phase 2 then also phase 3 is required.
The number of phases and their duration is assigned to a patient during the registration of the appointment.
Moreover, some patient might not need a bed or a chair, so for those patients the duration of phase 4 is 0.

Office hours are from 07:30 A.M. to 01:30 P.M. We consider a set of 72 time slots for each day with a duration of 5 minutes, where 07:30-07:35 is the first time slot, and 01:25-01:30 is the last one.
We also identify a set of 36 available time slots which represent the possible starting time of phase 4, where 07:35-07:40 is the first time slot, 07:45-07:50 is the second time slot, and so forth.
Moreover, if the duration of phase 4 for a registration exceeds a given threshold, then it must start after 11:25 A.M. (i.e., the 24th time slot). Finally, phase 4 must start after all previous phases are completed. 

The input of the problem consists of registrations, where each registration includes the duration of each phase (set to 0 if a phase is not required) and the preference between chair or bed.

A solution to the problem is represented by a schedule of registrations to time slots for a given day (representing the beginning of phase 4) according to the following requirements:
    \textit{(i)} each patient must be assigned to a chair or bed;
    \textit{(ii)} each chair or bed can be used by only one patient for each time slot; and
    \textit{(iii)} if the treatment requires more than one time slots, then the patient must always use the same chair or bed.
Moreover, an optimal solution to the problem maximizes the number of patients that are assigned to the preferred resource (chair or bed).
Ties are broken by minimizing the number of concurrent patients in phase 2 to have a more uniform usage of resources during the day.

\paragraph{Weekly CTS problem.}
The CTS problem aims at computing a scheduling solution for one day.
Nevertheless, the definition of the problem can be slightly modified for computing a scheduling solution for one week.
The main difference is that the registration includes a natural number expressing an ordering on the therapies for the patient.
For instance, given a patient, there might be several registrations associated to her/him which have different duration of the phases and different preferences between chair and bed.
Moreover, each patient must wait a certain number of days (depending on the therapy) from one appointment to the subsequent one.

\subsection{Formalization of the CTS problem}
\begin{definition}[Weekly CTS problem]
Let
\begin{itemize}
    \item $I$ be a finite set of identifiers of patients;
    \item $O \subset \mathbb{N}$ be a finite set of orders;
    \item $R \subseteq I \times O$ be a set of registrations, such that if $(i,o_1)\in R$ then for each $o_2 < o_1$, $(i,o_2) \in R$;
    \item $D$ be a finite set of days;
    \item $\mathit{ATS}=\{t \mid t \in [1..72]\}$ be the set of all time slots;
    \item $\mathit{TS}=\{t \cdot 2 \mid t \in [1..36]\}$ be the set of available time slots;
    \item $P=\{1, 2, 3, 4\}$ be the set of phases;
    \item $B=\{b_1, \ldots, b_m\}$ be a set of $m$ beds;
    \item $C=\{c_1, \ldots, c_n\}$ be a set of $n$ chairs;
    \item $\delta: R \times P \mapsto \mathbb{N}$ be a function associating a registration and a phase to a duration such that for a registration $r$ if $\delta(r, 2) \neq 0$ then $\delta(r,3) \neq 0$; 
    \item $\rho: R \mapsto \{\mathit{bed}, \mathit{chair}\}$ be a function associating the preference of a registration to beds or chairs.
    \item $\omega: R \mapsto \mathbb{N}^+$ be a function associating a registration to a waiting time.
\end{itemize}

Let $x:R \times D \times \mathit{TS} \times (B \cup C) \mapsto \{0, 1\}$ be a function such that $x(r, d, ts, s)=1$ if the registration $r$ is assigned to the day $d$ and time slot $ts$ with a bed or a chair $s$, and $0$ otherwise.
Moreover, for a given $x$ let $A_x = \{(r, d, ts, s) \mid r \in R, d \in D, ts \in \mathit{TS}, s \in (B \cup C), x(r,d, ts, s) = 1\}$.

Then, given sets $I$, $O$, $R$, $D$, $\mathit{ATS}$, $\mathit{TS}$, $P$, $B$, $C$, and functions $\delta$, $\rho$, $\omega$, the weekly CTS problem is defined as the problem of finding a schedule $x$, such that 
\begin{enumerate}[($c1$)]\small
    \item $|\{(d,ts) : (r,d,ts,s) \in A_x\}| = 1 \hfill \forall r \in R$; \label{cond:onereg}
    \item $|\{s : (r, d, ts, s) \in A_x\}| = 1 \hfill \forall r \in R \mid \delta(r,4) > 0$; \label{cond:samebedchair}
    \item $x(r_2, d, ts_2, s) = 0 \hfill \forall r_2 \in R, ts_2 \in \mathit{TS}, (r, d, ts, s) \in A_x \mid \ r \neq r_2, ts \leq ts_2 \leq ts + \delta(r, 4)$; \label{cond:onebedchair}
    \item $x(r,d,ts,s) = 0 \mid ts\in\{1,\ldots,23\} \hfill \forall r \in R, \delta(r,4) > 50$; \label{cond:longreg}
    \item $ts - \delta(r,1) - \delta(r,2) - \delta(r,3) > 0\hfill \forall (r, d, ts, s) \in A_x$; \label{cond:starttime}
    \item $|\{((i, o_1),d_1,ts_1,s_1) \in A_x\}| = 1 \hfill \forall ((i, o_2),d_2,ts_2,s_2) \in A_x \mid o_1=o_2+1,   d_1 = d_2+\omega(i,o_1)$. \label{cond:waiting}
\end{enumerate}
\end{definition}

Condition ($c\ref{cond:onereg}$) ensures that each registration must be assigned exactly once.
Condition ($c\ref{cond:samebedchair}$) ensures that each patient requiring a seat must be assigned to exactly one chair or bed.
Condition ($c\ref{cond:onebedchair}$) ensures that each chair or bed cannot be assigned to different patients during the same time slots.
Condition ($c\ref{cond:longreg}$) ensures that if phase 4 is particularly long, then it must start after 11:25 A.M. (i.e., the 25th time slot).
Condition ($c\ref{cond:starttime}$) ensures that phase 4 cannot start before other phases are concluded.
Condition ($c\ref{cond:waiting}$) ensures that each patient waits a given number of days between one appointment and the subsequent one.

\begin{definition}[Missed preferences]\label{def:2}
Given a solution $x$, let $m^*_x = |\{r \mid (r, d, ts, s) \in A_x, s \in B, \rho(r) = \mathit{chair}\} \cup \{r \mid (r, d, ts, s) \in A_x, s \in C, \rho(r) = \mathit{bed}\}|$. Intuitively, $m^*_x$ represents the number of registrations whose preference for beds/chairs are not fulfilled.
\end{definition}

\begin{definition}[Distribution of registrations]
Given a solution $x$ and a time slot $ts \in \mathit{ATS}$, let $d^{ts}_x = |\{r \mid (r, d, ts', s) \in A_x, \delta(r,2) \neq 0, ts=ts'-\delta(r,3)-\delta(r,2)\}|$, i.e., the number of registrations whose phase 2 starts at the time slot $ts$.
Given a solution $x$ and a day $d \in D$, let $g^d_x = |\{r \mid (r, d, ts, s) \in A_x\}|$, i.e., the number of registrations of the day $d$.
Given a solution $x$, let $d^*_x = \max(\{d^{ts}_x \mid ts \in \mathit{ATS}\}) - \min(\{d^{ts}_x \mid ts \in \mathit{ATS}\})$ and $g^*_x=\max(\{g^d_x \mid d \in D\})$.
\end{definition}

\begin{definition}[Optimal solution]\label{def:4}
A solution $x$ is said to dominate solution $x'$ if $m^*_x < m^*_{x'}$, or if $m^*_x = m^*_{x'}$ and $d^*_x < d^*_{x'}$, or if $m^*_x = m^*_{x'}$ and $d^*_x = d^*_{x'}$ and $g^*_x < g^*_{x'}$.
A solution is \textit{optimal} if it is not dominated by any other solution.
\end{definition}

\paragraph{Daily CTS problem.}
Finally, the daily CTS problem can be defined by setting $O=\{0\}$ and $D=\{1\}$, and, thus, Definition \ref{def:2}-\ref{def:4} also hold for it.

\section{ASP Encoding}
\label{sec:enc}
In this section we first present the ASP encoding for the weekly CTS problem, and then the main elements of the extended CTS problem, in two separate sub-sections.

\subsection{ASP Encoding for the (weekly) CTS problem}
\label{sec:modelsA}
\begin{figure}[t!]
\figrule
\begin{asp}
{x(RID,DAY,TS,PH4,0,S) : ts(TS), day(DAY)} = 1 :- reg(RID,0,_,PH4,PH3,PH2,PH1,S).
{x(RID,DAY+DAY2,TS,PH4,ORDER,S) : ts(TS)} = 1 :- x(RID,DAY,_,_,N,_), ORDER=N+1, day(DAY+DAY2), reg(RID,ORDER,DAY2,PH4,PH3,PH2,PH1,S).
:- x(RID,DAY,TS,PH4,_,_), PH4 > 50, TS < 24.
:- x(RID,DAY,TS,_,ORDER,_), reg(RID,ORDER,DAY2,PH4,PH3,PH2,PH1,S), TS-PH3-PH2-PH1<1.
1 {bed(ID,RID,DAY) : bed(ID); chair(ID,RID,DAY) : chair(ID)} 1:- x(RID,DAY,_,_,_,_).
res(RID,DAY,TS..TS+PH4-1) :- x(RID,DAY,TS,PH4,_,_), PH4 > 0.
chair(ID,RID,DAY,TS) :- chair(ID,RID,DAY), res(RID,DAY,TS).
bed(ID,RID,DAY,TS) :- bed(ID,RID,DAY), res(RID,DAY,TS).
:- #count{RID: chair(ID,RID,DAY,TS)} > 1, day(DAY), ts(TS), chair(ID).
:- #count{RID: bed(ID,RID,DAY,TS)} > 1, day(DAY), ts(TS), bed(ID).
support(RID,DAY,TS) :- x(RID,DAY,PH4,_,_,_), reg(RID,ORDER,_,_,PH3,PH2,_,_), PH2 > 0, TS=PH4-PH3-PH2, day(DAY), ats(TS).
numbReg(DAY,N,TS) :- N = #count{RID: support(RID,DAY,TS)}, day(DAY), ats(TS).
numMax(DAY,T) :- T = #max{N: numbReg(DAY,N,_)}, day(DAY).
numMin(DAY,T) :- T = #min{N: numbReg(DAY,N,_), N != 0}, day(DAY).
numbDay(DAY,N) :- N = #count{RID: support(RID,DAY,_)}, day(DAY).
numMaxDay(T) :- T = #max{N: numbDay(DAY, N)}.
:~ x(RID,DAY,_,_,_,"bed"), chair(ID,RID,DAY,_). [1@7,RID]
:~ x(RID,DAY,_,_,_,"chair"), bed(ID,RID,DAY,_). [1@7,RID]
:~ numMax(DAY,T). [T@6, DAY]
:~ numMax(DAY,MAX), numMin(DAY,MIN). [MAX-MIN@5,DAY]
:~ numMaxDay(N). [N@4]

\end{asp}
    \caption{ASP encoding of the real world problem}
    \label{fig:encoding1}
\figrule    
\end{figure}

We assume the reader is familiar with syntax and semantics of ASP.
Starting from the specifications in the previous section,
here we present the ASP encoding, based on the input language of {\sc clingo} \cite{DBLP:conf/iclp/GebserKKOSW16}. For details about syntax and semantics of ASP programs we refer the reader to the paper by \citeN{CalimeriFGIKKLM20}.

\paragraph{Data Model.}
The input data is specified by means of the following atoms:
\begin{itemize}
	\item Instances of \texttt{reg(RID,ORDER,WDAY,PH4,PH3,PH2,PH1,S)} represent the registrations, characterized by an id (\texttt{RID}), the ordering (\texttt{ORDER}), the number of waiting days before the visit (\texttt{WDAY}), the duration of each phase (\texttt{PH4}, ..., \texttt{PH1}), and the preference for chair or bed (\texttt{S}), where \texttt{S} can be \texttt{"bed"} or \texttt{"chair"}.
	\item Instances of \texttt{day(DAY)} represent the available days.
	\item Instances of \texttt{ts(TS)} represent the available time slots for phase 4, where \texttt{TS} ranges from \texttt{1} to \texttt{36}.
	\item Instances of \texttt{ats(TS)} represent all time slots, where \texttt{TS} ranges from \texttt{1} to \texttt{72}.
	\item Instances of \texttt{chair(ID)} represent the available chairs, with its identifier \texttt{ID}.
	\item Instances of \texttt{bed(ID)} represent the available beds, with its identifier \texttt{ID}.
\end{itemize}

The output is an assignment represented by atoms of the form:
\begin{equation*}
\texttt{x(RID,DAY,TS,PH4,ORDER,S)}    
\end{equation*}
where the intuitive meaning is that the phase 4 of registration with id \texttt{RID} and ordering \texttt{ORD} is assigned to the day \texttt{DAY} and time slot \texttt{TS} using the chair or bed \texttt{S}.

\paragraph{Encoding.}
The related encoding is shown in Figure~\ref{fig:encoding1}, and is described in the following.
To simplify the description, we denote as $r_i$ the rule appearing at line $i$ of Figure~\ref{fig:encoding1}.

Rule $r_1$ assigns registrations to a day and a time slot. The assignment is made only for the first registration of the patient (i.e., the one with ordering equal to 0).
Rule $r_2$ assigns subsequent registrations (i.e., the one with ordering greater than 0) to a time slot, whereas the day is automatically computed considering the number of waiting days before the visit.
Rules $r_3$ and $r_4$ encode conditions (c\ref{cond:longreg}) and (c\ref{cond:starttime}), respectively.  
Then, rule $r_5$ assigns exactly one chair/bed to each registration.
Rules from $r_6$ to $r_{10}$ are used to ensure that each chair and bed is assigned to at most one patient for each time slot.
Rules from $r_{12}$ to $r_{16}$ are needed to derive auxiliary atoms that are used later on in optimization.
In particular, rules from $r_{12}$ to $r_{14}$ compute the maximum and minimum number of patients that are simultaneously assigned to  phase 2, while rules $r_{15}$ and $r_{16}$ compute the number of patients for each day.

Finally, weak constraints $r_{17}$ and $r_{18}$ are used to minimize the number of missed preferences, whereas weak constraints from $r_{19}$ to $r_{21}$ enforce the distribution of registrations.

Note that the encoding can be already used for the daily CTS problem by considering only atoms of the form \texttt{reg(REGID,0,WDAY,PH4,PH3,PH2,PH1,S)} and only one day, e.g., \texttt{day(1)}.

\subsection{Extended CTS}\label{sec:extendedcts}
In this section we show how the daily CTS problem can be extended to deal with further requirements that were not considered by the solution of the San Martino Hospital of Genova.
The extensions are related to \textit{(a)} the drugs to be dispensed, and a daily limited availability for each drug; \textit{(b)} the priority level (high, medium, low) of the registration; and \textit{(c)} the number of nurses working in the hospital, where each nurse can assist at most $k$ patients per time slot.

Thus, in addition to the requirements defined in Section~\ref{sec:spec}, the solution scheduling has to guarantee that
\textit{(i)} each patient is assisted by exactly one nurse \cite{turkanscheduling2010};
\textit{(ii)} each nurse can assist from 1 to $k$ patients for each time slot \cite{turkanscheduling2010};
\textit{(iii)} treatments cannot exceed the maximum quantity of drugs available for each day \cite{Heshmat2021};
\textit{(iv)} treatments cannot be scheduled at the latest available time slot, since some drugs might require a long time to be prepared \cite{huggins_2014}; and 
\textit{(v)} registrations with the highest priorities should be scheduled before other registrations \cite{DBLP:conf/aiia/DodaroGMP18}.

\begin{figure}[t!]
\figrule
\begin{asp22}
reg(REGID,ORD,WDAY,PH4,PH3,PH2,PH1,S) :- reg(REGID,ORD,WDAY,PH4,PH3,PH2,PH1,S,PR,DRUG).
{nurses(ID,RID,DAY) : nurse(ID)} = 1 :- x(RID,DAY,_,_,_,_).
nurses(ID,RID,DAY,TS) :- nurses(ID,RID,DAY), res(RID,DAY,TS).
:- #count{RID: nurses(ID,RID,DAY,TS)} > K, day(DAY), ts(TS), nurse(ID), nurseLimits(K).
:- drug(DRUG,LMT,DAY), #count{RID: x(RID,DAY,_,_,ORDER,_), reg(RID,_,ORDER,_,_,_,_,_,_,DRUG)} > LMT.
:- x(RID,DAY,36,_,PH4,_,_).
:~ reg(RID,0,_,_,_,_,_,_,1,_), x(RID,_,TS,_,_,_). [TS@3,RID]
:~ reg(RID,0,_,_,_,_,_,_,2,_), x(RID,_,TS,_,_,_). [TS@2,RID]
:~ reg(RID,0,_,_,_,_,_,_,3,_), x(RID,_,TS,_,_,_). [TS@1,RID]
\end{asp22}
    \caption{ASP encoding of the added rules to the CTS extended problem}
    \label{fig:extended}
\figrule
\end{figure}

\paragraph{Data Model.}
For the extended problem we start from the atoms of the CTS problem, while changing the reg atom, and add further atoms:
\begin{itemize}
	\item Instances of \texttt{reg(REGID,ORD,WDAY,PH4,PH3,PH2,PH1,S,P,DR)} represent the registrations as in the CTS problem with the addition of a priority (\texttt{P}), and the drug (\texttt{DR}) required.
	\item Instances of \texttt{nurse(ID)} represent the available nurses, with its identifier \texttt{ID}.
	\item Instance of \texttt{nurseLimits(K)} represent the maximum number of patients (\texttt{K}) that a nurse can assist for each time slot.
	\item Instances of \texttt{drug(DR,LMT,DAY)} represent the amount of available drug for a given day, with its identifier \texttt{DR}, the available quantity \texttt{LMT}, and the day \texttt{DAY}.
\end{itemize}

The output is the same of the real world CTS problem.

\paragraph{Encoding.}
The encoding consists of the rules reported in Figure~\ref{fig:encoding1} (rules $r_1$--$r_{21}$) and the ones reported in Figure~\ref{fig:extended} (rules $r_{22}$--$r_{30}$).
Rule $r_{22}$ is needed to guarantee the backward compatibility with the previous encoding.
Rule $r_{23}$ assigns exactly one nurse to each patient (condition $(i)$).
Rules $r_{24}$ and $r_{25}$ encode condition $(ii)$, whereas $r_{26}$ and $r_{27}$ encode conditions $(iii)$ and $(iv)$, respectively.
Then, rules from $r_{28}$ to $r_{30}$ minimize the sum of time slots assigned to patients with priority levels equal to 1, 2, and 3, respectively.

\section{Experimental Results}\label{sec:experiments}
In this section we report the results of an empirical analysis of the CTS problem. Data are real world data from the San Martino Hospital for the CTS problem, while they have been randomly generated using parameters inspired by literature and real world data for the extended problem (e.g., from the work of \citeN{DBLP:conf/aiia/DodaroGMP18} for the distribution of priorities, and from the Oncology Nurses Society for the number of nurses per patient).
The generation of synthetic data for the extended problem is needed since real world data from the San Martino Hospital do not cover all the features, e.g., nurses and priorities, that we considered in the extended problem.
 
In this way we can simulate different scenarios and use them to test our encodings. The experiments were run on a AMD Ryzen 5 2600 CPU @ 3.40GHz with 15.9 GB of physical RAM. 
The ASP system used was \clingo~\cite{DBLP:conf/iclp/GebserKKOSW16} 5.1.0
, using parameters \textit{-{}-restart-on-model} for faster optimization and \textit{-{}-parallel-mode 8} for parallel execution. This setting is the result of a preliminary analysis done on the daily real world instances where we tested also other parameters, e.g., \texttt{--opt-strategy=usc} for optimization. The time limit was set to 60 seconds for the daily CTS and the extended problem, whereas for the weekly CTS was set to 1200 seconds. 
All material used in the experiments can be found at:
\url{http://www.star.dist.unige.it/~marco/ICLP2021/material.zip}.

\subsection{CTS benchmarks}\label{subsec:schedulerresults}
For the daily CTS we considered real data distributed in 22 days (two spare days of a week, Thursday and Friday, plus 4 consecutive weeks with working days Monday to Friday), each having an average number of 126 patients (12 std) with a minimum of 105 and a maximum of 148 patients per day.
Data related to the patients are taken from S. Martino Hospital and differ from each other in phases duration, phase 2 need and chair or bed request.
The duration of phase 4, which determines the duration of the occupation of the chair/bed, varies a lot from patient to patient. The majority of patients have a duration between 4 and 30 time slots but there are some outlier that require more than 72 time slots, i.e., these patients will end the treatment in the next session. 
Phase 2, that is the phase involved in the optimization, is requested by 44\% of the patients while 71\% of the patients require a chair and the others require a bed.

While for the real world (daily and weekly) problems we did not need to create synthetic data, for the extended one we need to assign a priority level to each patient, impose an availability for each drug used, define the number of nurses and the number of patients that a nurse can assist in the same slot.
The priorities of the registrations have been generated from an uneven distribution of three possible values (with weights respectively of 0.20, 0.40, and 0.40 for registrations having priority 1, 2, and 3, respectively).
In order to limit the degrees of our analysis, we decided to focus on nurses availability, and not limiting the drugs availability instead, and defined three scenario: one with low availability of nurses, one with an average availability and one having high availability. This choice is motivated by the real world scenario in the S. Martino Hospital, in which they would desire to limit and treat nurses explicitly, while quantity of drugs is not an issue.
To modify the availability we set the number of nurses to 5 and then changed the number of concurrent patients that can be assigned to the same nurse to 4, 7, and 10 for the three scenario, respectively. 

For the weekly problem, we considered the 4 weeks Monday to Friday and not considered the first two spare days, assigning a different regimen for the patients requiring more than one appointment.
For each week there are on average 634 registrations (26 std). Each registration is linked to a treatment of a patient, and the number of patients is lower than the total number of registrations (we remind that each patient can have more registrations). On average there are 542 patients for each week and the majority requires just one treatment, while about a 10\% require 2 or more treatments. 
For each patient, there could be up to 5 registrations, each corresponding to a day of treatment, following  different regimen. 

\subsection{Results for the real world problem}
\begin{figure}[t!]
\figrule
    \centering
\begin{tikzpicture}[scale=0.58]
\begin{axis}[ybar,
    width=1\textwidth,
	height=0.7\textwidth,
	font=\large,
	x = 0.4cm,
	enlarge x limits=0.02,
    ymin=0,
    ymax=10,        
    ylabel={Patients},
    major tick length=1pt, 
	xtick=data,
    xticklabels = {
        07:40, 07:50, 08:00, 08:10, 08:20, 08:30, 08:40, 08:50, 09:00, 09:10, 09:20, 09:30, 09:40, 09:50, 10:00, 10:10, 10:20, 10:30, 10:40, 10:50, 11:00, 11:10, 11:20, 11:30, 11:40, 11:50
    },
    x tick label style={rotate=90},
    bar width=0.1cm,
    title={(a) Best performance}
]
\addplot[draw=black,fill=blue] table[x index=0,y index=3] \dataset;
\addlegendentry{S. Martino scheduling}
\addplot[draw=black,fill=red] table[x index=0,y index=4] \dataset;
\addlegendentry{ASP scheduling}
\end{axis}
 \end{tikzpicture}
\begin{tikzpicture}[scale=0.58]
\begin{axis}[ybar,
    font=\large,
    width=1\textwidth,
	height=0.7\textwidth,
	x = 0.4cm,
	enlarge x limits=0.02,
    ymin=0,
    ymax=10,        
    ylabel={},
    major tick length=1pt, 
	xtick=data,
    xticklabels = {
        07:40, 07:50, 08:00, 08:10, 08:20, 08:30, 08:40, 08:50, 09:00, 09:10, 09:20, 09:30, 09:40, 09:50, 10:00, 10:10, 10:20, 10:30, 10:40, 10:50, 11:00, 11:10, 11:20, 11:30, 11:40, 11:50
    },
    x tick label style={rotate=90},
    bar width=0.1cm,
    title={(b) Worst performance}
]
\addplot[draw=black,fill=blue] table[x index=0,y index=1] \dataset;
\addlegendentry{S. Martino scheduling}
\addplot[draw=black,fill=red] table[x index=0,y index=2] \dataset;
\addlegendentry{ASP scheduling}
\end{axis}
 \end{tikzpicture}
\caption{\label{fig:compare} Number of patients assigned to their phase 2 for each time slot in a day, comparing our scheduling with the one of the S. Martino Hospital. Figure (a) [resp. (b)]   compares our best [resp. worst] result with the one of the Hospital in the same day.}
\figrule
\begin{tikzpicture}[scale=0.30]
\begin{axis}[scale only axis,
    width=1\textwidth,
	height=0.7\textwidth,
	font=\Huge,
	x = 0.3cm,
	enlarge x limits=0.02,
    ymin=0,
    ymax=6,        
    ylabel={Patients},
    major tick length=1pt, 
	xtick=data,
    xticklabels = {},
    x tick label style={rotate=90},
    title={Day 1}
]
\addplot [mark size=3.5pt, color=black, mark=o] table[x index=0,y index=1] \weeklydataset;
\end{axis}
 \end{tikzpicture}
\begin{tikzpicture}[scale=0.30]
\begin{axis}[scale only axis,
    width=1\textwidth,
	height=0.7\textwidth,
	font=\Huge,
	x = 0.3cm,
	enlarge x limits=0.02,
    ymin=0,
    ymax=6,        
    ylabel={},
    major tick length=1pt, 
	xtick=data,
    xticklabels = {},
    x tick label style={rotate=90},
    title={Day 2}
]
\addplot [mark size=3.5pt, color=black, mark=o] table[x index=0,y index=2] \weeklydataset;
\end{axis}
 \end{tikzpicture}
\begin{tikzpicture}[scale=0.30]
\begin{axis}[scale only axis,
    width=1\textwidth,
	height=0.7\textwidth,
	font=\Huge,
	x = 0.3cm,
	enlarge x limits=0.02,
    ymin=0,
    ymax=6,        
    ylabel={},
    major tick length=1pt, 
	xtick=data,
    xticklabels = {},
    x tick label style={rotate=90},
    title={Day 3}
]
\addplot [mark size=3.5pt, color=black, mark=o] table[x index=0,y index=3] \weeklydataset;
\end{axis}
 \end{tikzpicture}
\begin{tikzpicture}[scale=0.30]
\begin{axis}[scale only axis,
    width=1\textwidth,
	height=0.7\textwidth,
	font=\Huge,
	x = 0.3cm,
	enlarge x limits=0.02,
    ymin=0,
    ymax=6,        
    ylabel={},
    major tick length=1pt, 
	xtick=data,
    xticklabels = {},
    x tick label style={rotate=90},
    title={Day 4}
]
\addplot [mark size=3.5pt, color=black, mark=o] table[x index=0,y index=4] \weeklydataset;
\end{axis}
 \end{tikzpicture}
\begin{tikzpicture}[scale=0.30]
\begin{axis}[scale only axis,
    width=1\textwidth,
	height=0.7\textwidth,
	font=\Huge,
	x = 0.3cm,
	enlarge x limits=0.02,
    ymin=0,
    ymax=6,        
    ylabel={},
    major tick length=1pt, 
	xtick=data,
    xticklabels = {},
    x tick label style={rotate=90},
    title={Day 5}
]
\addplot [mark size=3.5pt, color=black, mark=o] table[x index=0,y index=5] \weeklydataset;
\end{axis}
 \end{tikzpicture}
\caption{\label{fig:compareweekly} Number of patients assigned to their phase 2 for each time slot and day during one working week.}
\figrule
\end{figure}
The first optimization criteria in the CTS problem is to assign the preferred resource (bed or chair) to as many patients as possible. Our solution is optimal in this respect, since it was able to accommodate \textit{all} preferences.
Moreover, we compared our results with the solution applied to the San Martino Hospital.
First of all, we noted that their solution needs to resort to \emph{virtual chairs}, meaning that some patients were not assigned to any bed nor chair. This was done to assign a time slot to all the patients, even if it implies that there might be time slots with more patients than those treatable in each time slot with the available resources.
Our solution does not require such virtual chairs, i.e., all the patients are assigned to an available bed or chair, and it assigns a resource (bed or chair) to 207 more patients than the one of the Hospital.

The second optimization criteria is to have a more equally distributed affluence of patients during their phase 2 for each time slot.
Results are reported in Figure~\ref{fig:compare},  which details the number of patients starting phase 2 in each time slot; the figure compares our best (a) and worst (b) results with those of the Hospital in the same days.
From the graphs it is clear that even our worst performance produces highly balanced results.
It is important to emphasize here that during the search for the optimal solution \clingo produces several suboptimal ones, i.e., it has an \textit{anytime behavior}. In our case, we observed that \clingo was able to compute the optimal solution even if it could not prove its optimality within the time limit.

Concerning the weekly problem, also in this case our solution was always able to fulfill the preferences of the patients concerning bed or chairs, while for the second optimization criteria results are slightly worse than the ones for the daily problem.
In particular, Figure~\ref{fig:compareweekly} reports the number of patients for each hour and for each day starting phase 2 of one week (results of other weeks are similar). Even if in the first two days the results are worse, as expected, compared to the results on one day, they are still better than the ones implemented by S. Martino, while for days 4 and 5 they are as good as the one for the analysis on the single day. Thus, overall our solution proved to be a viable tool to schedule the all week, considering that being able to schedule the whole week in advance could lead to benefits especially from an organizational point of view.

\paragraph{Comparison to alternative logic-based formalisms.}
In the following, we present an empirical comparison of our ASP-based solution with alternative logic-based approaches, obtained by applying automatic translations of ASP instances.
In more detail, we used the ASP solver \textsc{wasp}~\cite{DBLP:conf/lpnmr/AlvianoADLMR19}, with the option \texttt{--pre=wbo}, which converts ground ASP instances into pseudo-Boolean instances in the wbo format~\cite{pbcompetition}.
Then, we used the tool \textsc{pypblib}~\cite{pypblib} to encode wbo instances as MaxSAT instances.
Moreover, in order to provide a fair comparison, we also processed our ASP instances using \textsc{wasp} with the option \texttt{--pre=lparse}, which collapses all weak constraints levels into one single level using exponential weights. In this way, the costs found by the different approaches can be compared.

Then, we considered three state-of-the-art MaxSAT solvers, namely \textsc{MaxHS}~\cite{davies2013solving}, \textsc{open-wbo}~\cite{DBLP:conf/sat/MartinsML14}, and \textsc{rc2}~\cite{DBLP:journals/jsat/IgnatievMM19}, and the industrial tool for solving optimization problems \textsc{gurobi}~\cite{gurobi}, which is able to process instances in the wbo format.
Concerning \textsc{clingo}, we used \textit{(i)} its default configuration (\textsc{clingo-def}); \textit{(ii)} the option \texttt{restart-on-model} (\textsc{clingo-rom}); and \texttt{(iii)} the option \texttt{--opt-strategy=usc} (\textsc{clingo-usc}). The latter enables the usage of algorithm \textsc{oll}~\cite{DBLP:conf/cp/MorgadoDM14}, which is the same algorithm employed by the MaxSAT solver \textsc{rc2}.

\begin{table}[t]\footnotesize
    \caption{Comparison of ASP solution with alternative logic-based solutions.}
 	\centering
 	 \label{tab:resultscomparisonotherformalisms}
 	\begin{tabular}{cccccccc}
 		\hline\hline
Instance & \textsc{clingo-def} & \textsc{clingo-rom} & \textsc{clingo-usc} & \textsc{MaxHS} & \textsc{open-wbo} & \textsc{rc2} & \textsc{gurobi} \\
\cmidrule{1-1}\cmidrule{2-2} \cmidrule{3-3}\cmidrule{4-4} \cmidrule{5-5} \cmidrule{6-6} \cmidrule{7-7} \cmidrule{8-8}
10-01 & 1 & 1 & - & 4 & 3 & - & -\\
10-02 & 1 & 1 & - & 4 & 3 & - & -\\
10-05 & 1 & 2 & - & 4 & 3 & - & -\\
10-06 & 1 & 1 & - & 4 & 3 & - & -\\
10-07 & 1 & 1 & - & 4 & 3 & - & -\\
10-08 & 1 & 1 & - & 4 & 3 & - & -\\
10-09 & 2 & 1 & - & 4 & 3 & - & -\\
10-13 & 1 & 1 & - & 4 & 3 & - & -\\
10-14 & 3 & 1 & - & 4 & 2 & - & -\\
10-15 & 2 & 2 & - & 4 & 1 & - & -\\
10-16 & 2 & 1 & - & 4 & 2 & - & -\\
10-19 & 1 & 1 & - & 4 & 3 & - & -\\
10-20 & 1 & 1 & - & 4 & 3 & - & -\\
10-21 & 1 & 1 & - & 4 & 3 & - & -\\
10-22 & 1 & 1 & - & 4 & 3 & - & -\\
10-23 & 1 & 1 & - & 3 & 4 & - & -\\
10-26 & 2 & 1 & - & 4 & 3 & - & -\\
10-27 & 1 & 1 & - & 4 & 3 & - & -\\
10-28  & 1 & 1 & - & 4 & 1 & - & -\\
10-29 & 1 & 1 & - & 4 & 3 & - & -\\
10-30 & 1 & 1 & - & 4 & 3 & - & -\\
        \hline\hline
    \end{tabular}
\end{table}

The experiment was executed on the daily instances, with a timeout of 60 seconds as in Section~\ref{subsec:schedulerresults}.
Results are reported in Table~\ref{tab:resultscomparisonotherformalisms}, where for each solver and instance we report the ranking obtained by each solver. The solver is in the first position if it finds the solution with the lowest cost, a dash means that the solver outputs no solution within the time limit.
As a general observation, \textsc{clingo-rom} obtains the best performance overall, since it finds the best cost in all but two instances.
The performance of \textsc{clingo-def} is in general slightly worse than the one of \textsc{clingo-rom}, even if in the majority of the instances they compute the same solution.
Concerning MaxSAT solvers, we observe that \textsc{open-wbo} is the best performing solver, and it is able to produce the best solution on instances 10-15 and 10-28. \textsc{MaxHS} is able to print some solution within the time limit, which is however often much worse than the ones found by \textsc{clingo-rom}, \textsc{clingo-def}, and \textsc{open-wbo}. The only exception is represented by instance 10-23 where \textsc{clingo-rom} and \textsc{clingo-def} find a solution with a cost equal to 116489, and \textsc{MaxHS} finds a solution with a cost equal to 116490.
Concerning \textsc{clingo-usc} and \textsc{rc2}, we observe that both solvers are based on the same algorithm, namely \textsc{oll}~\cite{DBLP:conf/cp/MorgadoDM14}, which however is able to produce only optimal solutions and none is found within the time limit.
Finally, we mention that also \textsc{gurobi} cannot compute a solution within the time limit.


\subsection{Results for the extended problem}
First, we note that also in this case our solution is able to assign the preferred resource to each patient. Regarding uniformity of patients starting phase 2, Figure~\ref{fig:comparenurses} reports the results on our three scenario about nurses availability. Each graph is organized as Figure~\ref{fig:compare}. It is possible to observe that with low availability (left) nurses are not enough to satisfy the desired property of having a (approx.) constant number of patients during the session, while with medium (center) and high (right) availability results are very positive and similar to the daily CTS problem. It is important to note that we are still able to reach an optimal result in the same amount of time with high availability of nurses even with the extended case.

\begin{figure}[t!]
\figrule
    \centering
\begin{tikzpicture}[scale=0.4]
\begin{axis}[scale only axis,
    width=1\textwidth,
	height=0.7\textwidth,
	font=\Large,
	x = 0.4cm,
	enlarge x limits=0.02,
    ymin=0,
    ymax=10,        
    ylabel={Patients},
    major tick length=1pt, 
	xtick=data,
    xticklabels = {
        07:40, 07:50, 08:00, 08:10, 08:20, 08:30, 08:40, 08:50, 09:00, 09:10, 09:20, 09:30, 09:40, 09:50, 10:00, 10:10, 10:20, 10:30, 10:40, 10:50, 11:00, 11:10, 11:20, 11:30, 11:40, 11:50
    },
    x tick label style={rotate=90},
    bar width=0.1cm,
    title={Low availability}
]
\addplot [mark size=3.5pt, color=black, mark=o] table[x index=0,y index=1] \extendedctsdataset;
\end{axis}
 \end{tikzpicture}
\begin{tikzpicture}[scale=0.4]
\begin{axis}[scale only axis,
    width=1\textwidth,
	height=0.7\textwidth,
	font=\Large,
	x = 0.4cm,
	enlarge x limits=0.02,
    ymin=0,
    ymax=10,        
    ylabel={},
    major tick length=1pt, 
	xtick=data,
    xticklabels = {
        07:40, 07:50, 08:00, 08:10, 08:20, 08:30, 08:40, 08:50, 09:00, 09:10, 09:20, 09:30, 09:40, 09:50, 10:00, 10:10, 10:20, 10:30, 10:40, 10:50, 11:00, 11:10, 11:20, 11:30, 11:40, 11:50
    },
    x tick label style={rotate=90},
    bar width=0.1cm,
    title={Medium availability}
]
\addplot [mark size=3.5pt, color=black, mark=o] table[x index=0,y index=2] \extendedctsdataset;
\end{axis}
 \end{tikzpicture}
\begin{tikzpicture}[scale=0.4]
\begin{axis}[scale only axis,
    width=1\textwidth,
	height=0.7\textwidth,
	font=\Large,
    x = 0.4cm,
	enlarge x limits=0.02,
    ymin=0,
    ymax=10,        
    ylabel={},
    major tick length=1pt, 
	xtick=data,
    xticklabels = {
        07:40, 07:50, 08:00, 08:10, 08:20, 08:30, 08:40, 08:50, 09:00, 09:10, 09:20, 09:30, 09:40, 09:50, 10:00, 10:10, 10:20, 10:30, 10:40, 10:50, 11:00, 11:10, 11:20, 11:30, 11:40, 11:50
    },
    x tick label style={rotate=90},
    bar width=0.1cm,
    title={High availability}
]
\addplot [mark size=3.5pt, color=black, mark=o] table[x index=0,y index=3] \extendedctsdataset;
\end{axis}
 \end{tikzpicture}
\caption{\label{fig:comparenurses} Number of patients assigned to their phase 2 in the extended CTS with (left) low, (center) medium and (right) high availability of nurses.}
\figrule
\begin{tikzpicture}[scale=0.35]
    \begin{axis}[
        ybar stacked,
        x label style = {at={(axis description cs:0.5,0.03)}},
        y label style = {at={(axis description cs:0.05,0.5)}},
        width=1\textwidth,
	    height=0.7\textwidth,
	    font=\large,
	    x = 0.5cm,
	    enlarge x limits=0.02,
        ymin=0,
        ymax=25,
        xtick=data,
        major tick length=1pt, 
        xticklabels = {
        07:40, 07:50, 08:00, 08:10, 08:20, 08:30, 08:40, 08:50, 09:00, 09:10, 09:20, 09:30, 09:40, 09:50, 10:00, 10:10, 10:20, 10:30, 10:40, 10:50, 11:00, 11:10, 11:20, 11:30, 11:40, 11:50, 12:00, 12:10, 12:20, 12:30, 12:40, 12:50, 13:00, 13:10, 13:20
    },
        x tick label style={rotate=90},
        bar width=0.2cm,
        title={(a) Low availability}
    ]
        \addplot [fill=blue!80]
            table [y index=1, x index=0]
                {\prioritytest};
                    \addlegendentry{Priority1}
        \addplot [fill=red!60]
            table [y index=2, x index=0]
                {\prioritytest};
                    \addlegendentry{Priority2}
        \addplot [fill=green!60]
            table [y index=3, x index=0]
                {\prioritytest};
                    \addlegendentry{Priority3}
    \end{axis}
\end{tikzpicture}
\begin{tikzpicture}[scale=0.35]
    \begin{axis}[
        ybar stacked,
        x label style = {at={(axis description cs:0.5,0.03)}},
        y label style = {at={(axis description cs:0.05,0.5)}},
        width=1\textwidth,
	    height=0.7\textwidth,
	    font=\large,
	    x = 0.5cm,
	    enlarge x limits=0.02,
        ymin=0,
        ymax=25,
        xtick=data,
        major tick length=1pt, 
        xticklabels = {
        07:40, 07:50, 08:00, 08:10, 08:20, 08:30, 08:40, 08:50, 09:00, 09:10, 09:20, 09:30, 09:40, 09:50, 10:00, 10:10, 10:20, 10:30, 10:40, 10:50, 11:00, 11:10, 11:20, 11:30, 11:40, 11:50, 12:00, 12:10, 12:20, 12:30, 12:40, 12:50, 13:00, 13:10, 13:20
    },
        x tick label style={rotate=90},
        bar width=0.2cm,
        title={(a) Medium availability}
    ]
        \addplot [fill=blue!80]
            table [y index=4, x index=0]
                {\prioritytest};
                    \addlegendentry{Priority1}
        \addplot [fill=red!60]
            table [y index=5, x index=0]
                {\prioritytest};
                    \addlegendentry{Priority2}
        \addplot [fill=green!60]
            table [y index=6, x index=0]
                {\prioritytest};
                    \addlegendentry{Priority3}
    \end{axis}
\end{tikzpicture}
\begin{tikzpicture}[scale=0.35]
    \begin{axis}[
        ybar stacked,
        x label style = {at={(axis description cs:0.5,0.03)}},
        y label style = {at={(axis description cs:0.05,0.5)}},
        width=1\textwidth,
	    height=0.7\textwidth,
	    font=\large,
	    x = 0.5cm,
	    enlarge x limits=0.02,
        ymin=0,
        ymax=25,
        xtick=data,
        major tick length=1pt, 
        xticklabels = {
        07:40, 07:50, 08:00, 08:10, 08:20, 08:30, 08:40, 08:50, 09:00, 09:10, 09:20, 09:30, 09:40, 09:50, 10:00, 10:10, 10:20, 10:30, 10:40, 10:50, 11:00, 11:10, 11:20, 11:30, 11:40, 11:50, 12:00, 12:10, 12:20, 12:30, 12:40, 12:50, 13:00, 13:10, 13:20
    },
        x tick label style={rotate=90},
        bar width=0.2cm,
        title={(a) High availability}
    ]
        \addplot [fill=blue!80]
            table [y index=7, x index=0]
                {\prioritytest};
                    \addlegendentry{Priority1}
        \addplot [fill=red!60]
            table [y index=8, x index=0]
                {\prioritytest};
                    \addlegendentry{Priority2}
        \addplot [fill=green!60]
            table [y index=9, x index=0]
                {\prioritytest};
                    \addlegendentry{Priority3}
    \end{axis}
\end{tikzpicture}
\caption{\label{fig:priority} Distribution of patients for the treatment and their priority in the extended problem with low (left), medium (center) and high availability (right) of nurses.}
\figrule
\end{figure}

The distribution of patients based on their priority in the time slots of a day for the three scenario is shown in Figure~\ref{fig:priority}, in which we can see that in the low availability scenario the distribution does not respect the priority adequately, while with medium and high availability the results are highly satisfying, through not optimal.
In the scenario with high availability of nurses, the majority of patients with high priority are scheduled in the first two available time slots, whereas in the latest time slots there are almost only patients with low priority.

It is finally interesting to note that such results, for our more involved setting, are reached despite the weak constraints related to the availability of nurses have lower level than the previous weak constraints, and the still low time limit (60 seconds). 

\section{Rescheduling}
\label{sec:resch}
In this section, we show our solution to a rescheduling situation, in which the original schedule cannot be fulfilled as planned.
It is important to emphasize here that rescheduling is usually independent from the quality of the schedule, and it is usually due to unpredictable events. 
First of all, even if every treatment has a regimen a doctor can try to personalize the treatment for a patient and try to change the regimen to have a better response.
Therefore, doctors should have the possibility to modify appointments for a patient without impacting too much the original calendar.
Then, a patient may be ill or unable to show up for a visit, and it is important to be able to give a new appointment as soon as possible.
Having a proper solution to these situations can improve the chance of survival of patients, since several recent studies \cite{sud_collateral_2020,kumar_treatment_2020} showed that delay of treatments, due to the COVID19 emergency, has had a negative impact on the survival rates of the patients.

In our setting, the rescheduling is applied to a weekly schedule, where a number of appointments are canceled, regimen are possibly changed, and thus they need to be rescheduled.
The generated output has to follow the same requirements of the weekly scheduling plus some additional requirements:
    \textit{(i)} changed appointments can only be postponed, i.e., appointments cannot be moved before the original day of the appointment because patients could be not ready;
    \textit{(ii)} no appointment can be changed before the first day in which a patient requires a modification to her/his registration or there is an unavailable patient; this is required to limit the changes to the calendar;
    \textit{(iii)} a patient that has already completed the first appointment but has neither unavailable days nor modified registrations cannot be postponed, to ensure compliance with the regimen; and
    \textit{(iv)} when the first appointment of a patient is postponed, then all subsequent appointments of the same patient should be rescheduled to try to follow the regimen.
    
Moreover, an optimal solution has to $(a)$ minimize the sum of the distance between the days required by a regimen and the actual day of the appointment, $(b)$ minimize the distance between the new day for the first appointment and the old one for the patients with a postponed appointment, while $(c)$ maximizing the number of patients that are assigned to the preferred resource.

\paragraph{Data Model.}


The input data is the same of the weekly problem plus the following atoms:

\begin{itemize}
	\item Instances of \texttt{x(RID,DAY,TS,DUR,ORDER,S)} represent the previous scheduling, i.e., the output of the encodings described in Section~\ref{sec:enc}.
	\item Instances of \texttt{un(RID, DAY)} represent the day (\texttt{DAY}) in which the patient characterized by an id (\texttt{RID}) is unavailable.
	\item Instances of \texttt{regN(REGID,ORD,WDAY,PH4,PH3,PH2,PH1,S)} represent the new registrations due to changes of regimens decided by doctors,
	characterized by an id (\texttt{REGID}), the ordering (\texttt{ORD}), the number of waiting days before the visit (\texttt{WDAY}), the duration of each phase (\texttt{PH4}, ..., \texttt{PH1}), and the preference for chair or bed (\texttt{C}), where \texttt{C} can be \texttt{"bed"} or \texttt{"chair"}. 
\end{itemize}

The output is similar to the encodings described in Section~\ref{sec:enc}, where the schedule is represented by atoms of the form \texttt{y(RID,DAY,TS,DUR,ORDER,S)}.

\paragraph{ASP Encoding.}
\label{sec:models}
\begin{figure}[t!]
\figrule
\begin{asp}
{y(RID,DAY,TS,PH4,0,C) : ts(TS), day(DAY), not un(RID,DAY)} = 1 :- reg(RID,0,_,PH4,PH3,PH2,PH1,S), not regN(RID,0,_,_,_,_,_,_).
{y(RID,DAY,TS,PH4,0,C) : ts(TS), day(DAY), not un(RID,DAY)} = 1 :- regN(RID,0,_,PH4,PH3,PH2,PH1,S).
y(RID,DAY,TS,PH4,ORDER,C) :- y(RID,DAY,_,_,N,_), x(RID,DAY,_,_,N,_), x(RID,DAY,TS,PH4,ORDER,C), not regN(RID,ORDER,_,_,_,_,_,_), not un(RID,DAY).   
{y(RID,DAY,TS,PH4,ORDER,C) : ts(TS), day(DAY), not un(RID,DAY)} = 1 :- y(RID,DAY1,_,_,N,_), reg(RID,ORDER,DAY2,PH4,PH3,PH2,PH1,S), not regN(RID,ORDER,_,_,_,_,_,_), ORDER=N+1, day(DAY1+DAY2).
{y(RID,DAY,TS,PH4,ORDER,C) : ts(TS), day(DAY), not un(RID,DAY)} = 1 :- y(RID,DAY1,_,_,N,_), regN(RID,ORDER,DAY2,PH4,PH3,PH2,PH1,S), ORDER=N+1, day(DAY1+DAY2).
y(RID,DAY,TS,PH4,ORDER,C) :- x(RID,DAY,TS,PH4,ORDER,C), un(RID,_), DAY < #min{D: un(REGID,D)}.
y(RID,DAY,TS,PH4,ORDER,C) :- x(RID,DAY,TS,PH4,ORDER,C), DAY < #min{D: un(_,D);D: regN(RID,ORDER,_,_,_,_,_,_), x(RID,D,_,_,_,ORDER,_)}.
:- y(RID,DAY1,_,_,ORDER,_), x(RID,DAY2,_,_,ORDER,_), not regN(RID,_,_,_,_,_,_,_), DAY1 < DAY2.
:~ y(REGID,DAY1,_,_,_,N,_), y(REGID,DAY2,_,_,_,ORDER,_), reg(RID,ORDER,DD,_,_,_,_,_), not regN(RID,ORDER,_,_,_,_,_,_), ORDER=N+1. [|DD - (DAY2 -DAY1|@8,RID,ORDER] 
:~ y(REGID,DAY1,_,_,_,N,_), y(REGID,DAY2,_,_,_,ORDER,_), not regN(RID,ORDER,ORDER,_,_,_,_,_), ORDER=N+1. [|DD - (DAY2 -DAY1|@8,RID,ORDER] 
:~ y(RID, DAY1,_,_,_,0,_), x(RID, DAY2,_,_,_,0,_). [|DAY1 - DAY2|@7,RID]
\end{asp}
    \caption{ASP encoding of the rescheduling problem}
    \label{fig:encoding2}
\figrule    
\end{figure}
The encoding for the rescheduling problem is made of the rules in Figure~\ref{fig:encoding2}, where again to simplify the description we denote as $r_i$ the rule appearing at the line $i$, plus rules from $r_3$ to $r_{10}$ and weak constraints $r_{17}$ and $r_{18}$ from Figure~\ref{fig:encoding1} where atom $y$ replaces atom $x$. 

Rule $r_1$ and $r_2$ assign registrations to a day and a time slot, where $r_1$ assigns a day and a time slot to patient without a new registration, while $r_2$ assigns a day and a time slot using new registrations. The assignment is made only for the first registration of the patient (i.e., the one where ordering is equal to 0). Rule $r_3$ ensures that assignment of the previous scheduling is preserved for patients that need no modification. 
Rule $r_4$ assigns a day and a time slot to patients without a new registration, while $r_5$ assigns a day and a time slot using new registrations, for the subsequent registrations following the first one.
Rule $r_6$ ensures that the assignments before the first day of unavailability of that patient are not changed.
Rule $r_7$ ensures that all the assignments are not changed before the first day in which a patient is unavailable or is required a new registration.
Then, rule $r_8$ ensures that all the changed assignments are not anticipated.
Finally, weak constraints $r_{9}$ and $r_{10}$ are used to minimize the sum of the distance between the days requested by a regimen and the actual day of the appointment assigned to a patient, considering old registrations in $r_9$ and the new registrations in $r_{10}$, whereas weak constraint $r_{11}$ minimizes the distance between the day assigned to the first assignment in the old and new schedules.

\paragraph{Experiments.}
We tested our solution to the rescheduling problem in a number of different scenarios.
Scenario I, II, and III consider a sudden unavailability of 15, 20, and 25 patients, respectively, whereas Scenario IV,V, and VI considers a sudden unavailability of 15 patients and 1, 2, and 3 regimens, respectively, are modified.
Results are summarized in Table~\ref{tab:reschedule}.
First of all, it is important to emphasize that all obtained solutions are optimal and that all of them were computed in less than 4 minutes.
Indeed, the ASP solution is able to reschedule the patients without changing any unnecessary registration, i.e., no patient has the first appointment changed (apart from the ones with unavailability). We also mention that our solutions always assign the preferred resource (bed or chair) to all the patients. We finally report that further scenario with 20 unavailable patients and changed regimen, not shown in the table, confirm the results.
\begin{table}[t]
    \caption{Summary of the rescheduling results.}
 	\centering
 	\label{tab:reschedule}
 	\begin{tabular}{ccccc}
 	    \hline
 		\hline
 		Scenario & Unavailable patients & Changed regimen & Unnecessary registration moved & Time \\ 
\cmidrule{1-1} \cmidrule{2-3} \cmidrule{4-5} 		
 		I    & 15   & -   & 0   & 196 s       \\
 		II   & 20   & -   & 0   & 218 s       \\
 		III  & 25   & -   & 0   & 229 s        \\
 		IV   & 15   & 1   & 0   & 190 s         \\
 		IV   & 15   & 2   & 0   & 203 s         \\
 		IV   & 15   & 3   & 0   & 205 s         \\
 		\hline\hline
 	\end{tabular}
\end{table}

\section{Related Work}
\label{sec:rel}
\citeN{sevinc_algorithms_2013} addressed the CTS problem through a two-phase approach. In the first one an adaptive negative-feedback scheduling algorithm is adopted to control the load on the system, while in the second phase two heuristics based on the ‘Multiple Knapsack Problem’ have been evaluated to assign patients to specific infusion seats. The overall design has been tested at a local chemotherapy center and has yielded good results for patient waiting times, orderly execution of chemotherapy regimen and utilization of infusion chairs. \citeN{huang_alternative_2017} developed and implemented a model to optimize safety and efficiency in terms of
staffing resource violations measured by nurse-to-patient ratios throughout the workday and at key points during treatment to decide when to schedule patients according to their visit duration. The optimization model was built using Excel Solver.
\citeN{hahn-goldberg_dynamic_2014} addressed in particular dynamic uncertainty that arises from requests for appointments that arrive in real time and uncertainty due to last minute scheduling changes through a proactive template of an expected day in the chemotherapy centre using a deterministic optimization model updated, to accommodate last minute additions and cancellations to the schedule, by a shuffling algorithm. \citeN{huggins_2014} presented a mixed-integer programming optimization model developed with the objective of maximizing resource utilization, while balancing human workload, in particular taking into account variability in length of treatment, increased patient demand, and resource limitations.
Compared to the papers mentioned above, our solution tackles further aspects.
For the scheduling problem, in addition to taking into account the starting time of treatments, we also manage the previous phases (blood collection and visit). 
We also minimize the number of patients who request blood collection at the same time.
Considering these previous phases is important because, as mentioned by \citeN{Lame_2016}, usually the CTS problem involves many departments and knowing when a resource will be used is vital for an efficient planning.
While in our work we focused on rescheduling as a tool that allows to face unexpected events, the rescheduling proposed by \citeN{hahn-goldberg_dynamic_2014} is implemented as a way to reach better results when new patients are added.
Another difference is that \citeN{hahn-goldberg_dynamic_2014} apply the rescheduling to a daily problem, while we consider a 5 days horizon and more registrations for each patient.

ASP has been already used as a tool for solving scheduling problems \cite{DBLP:journals/aim/ErdemGL16,DBLP:journals/tplp/GebserOSR18}, also in the health care domains~\cite{DBLP:conf/aiia/AlvianoBCDGKMMM20}. In this paper, we focus on a novel problem whose requirements, differently from previous work, were completely specified by the S. Martino Hospital, and we employed real world data for the empirical analysis.

Finally, we mention that this paper extends and revises a paper appearing in the informal CEUR Proceedings \cite{DodaroGMMP20}, with the following additions: \textit{(i)} a mathematical formulation of the problem, \textit{(ii)} encodings for a real world problem, \textit{(iii)} a rescheduling solution, and \textit{(iv)} all analysis performed on real data.

\section{Conclusions and Current Work}
\label{sec:conc}
In this paper, we have employed ASP for solving some variants of the CTS problem. We started from the problem addressed by the S. Martino Hospital in Genova, then considered longer planning horizons, and desired features not taken at the moment into account by the hospital. Results on real data show that our solution is able, for several of such variants, to balancing in a highly satisfiable way the number of patients during the days and/or the slots of a day, and in short time. We also presented a rescheduling solution that shows flexibility in dealing with unavailability and changed regimen, minimizing the number of changes to be done on the original schedule.
Future work includes investigating the possibility to use extensions of ASP, such as ASP with preferences (e.g., using \textsc{asprin}~\cite{DBLP:conf/aaai/BrewkaD0S15}), since it can be effective in solving multi-objective version of CTS.

\section*{Acknowledgments} The work is carried out in the framework of a partnership with Janssen-Cilag SpA which partially sponsored the research. Authors are also grateful to the staff working at the San Martino Hospital for their support and for the data used in the paper.

\bibliographystyle{acmtrans}
\bibliography{bibtex}

\begin{thebibliography}{}

\bibitem[\protect\citeauthoryear{Alviano, Amendola, Dodaro, Leone, Maratea, and
  Ricca}{Alviano et~al\mbox{.}}{2019}]{DBLP:conf/lpnmr/AlvianoADLMR19}
{\sc Alviano, M.}, {\sc Amendola, G.}, {\sc Dodaro, C.}, {\sc Leone, N.}, {\sc
  Maratea, M.}, {\sc and} {\sc Ricca, F.} 2019.
\newblock Evaluation of disjunctive programs in {WASP}.
\newblock In {\em {LPNMR} 2019}. LNCS, vol. 11481. Springer, 241--255.

\bibitem[\protect\citeauthoryear{Alviano, Bertolucci, Cardellini, Dodaro,
  Galat{\`{a}}, Khan, Maratea, Mochi, Morozan, Porro, and Schouten}{Alviano
  et~al\mbox{.}}{2020}]{DBLP:conf/aiia/AlvianoBCDGKMMM20}
{\sc Alviano, M.}, {\sc Bertolucci, R.}, {\sc Cardellini, M.}, {\sc Dodaro,
  C.}, {\sc Galat{\`{a}}, G.}, {\sc Khan, M.~K.}, {\sc Maratea, M.}, {\sc
  Mochi, M.}, {\sc Morozan, V.}, {\sc Porro, I.}, {\sc and} {\sc Schouten, M.}
  2020.
\newblock Answer set programming in healthcare: Extended overview.
\newblock In {\em IPS and RCRA 2020}. {CEUR} Workshop Proceedings, vol. 2745.
  CEUR-WS.org.

\bibitem[\protect\citeauthoryear{Ansótegui, Pacheco, and Pon}{Ansótegui
  et~al\mbox{.}}{2019}]{pypblib}
{\sc Ansótegui, C.}, {\sc Pacheco, T.}, {\sc and} {\sc Pon, J.} 2019.
\newblock Pypblib.

\bibitem[\protect\citeauthoryear{Brewka, Delgrande, Romero, and Schaub}{Brewka
  et~al\mbox{.}}{2015}]{DBLP:conf/aaai/BrewkaD0S15}
{\sc Brewka, G.}, {\sc Delgrande, J.~P.}, {\sc Romero, J.}, {\sc and} {\sc
  Schaub, T.} 2015.
\newblock asprin: Customizing answer set preferences without a headache.
\newblock In {\em {AAAI} 2015}. {AAAI} Press, 1467--1474.

\bibitem[\protect\citeauthoryear{Calimeri, Faber, Gebser, Ianni, Kaminski,
  Krennwallner, Leone, Maratea, Ricca, and Schaub}{Calimeri
  et~al\mbox{.}}{2020}]{CalimeriFGIKKLM20}
{\sc Calimeri, F.}, {\sc Faber, W.}, {\sc Gebser, M.}, {\sc Ianni, G.}, {\sc
  Kaminski, R.}, {\sc Krennwallner, T.}, {\sc Leone, N.}, {\sc Maratea, M.},
  {\sc Ricca, F.}, {\sc and} {\sc Schaub, T.} 2020.
\newblock Asp-core-2 input language format.
\newblock {\em Theory and Practice of Logic Programming\/}~{\em 20,\/}~2,
  294--309.

\bibitem[\protect\citeauthoryear{Davies}{Davies}{2013}]{davies2013solving}
{\sc Davies, J.} 2013.
\newblock Solving maxsat by decoupling optimization and satisfaction.
\newblock Ph.D. thesis, University of Toronto.

\bibitem[\protect\citeauthoryear{Dodaro, Galat{\`{a}}, Maratea, Mochi, and
  Porro}{Dodaro et~al\mbox{.}}{2020}]{DodaroGMMP20}
{\sc Dodaro, C.}, {\sc Galat{\`{a}}, G.}, {\sc Maratea, M.}, {\sc Mochi, M.},
  {\sc and} {\sc Porro, I.} 2020.
\newblock Chemotherapy treatment scheduling via answer set programming.
\newblock In {\em {CILC} 2020}. {CEUR} Workshop Proceedings, vol. 2710.
  CEUR-WS.org, 342--356.

\bibitem[\protect\citeauthoryear{Dodaro, Galat{\`{a}}, Maratea, and
  Porro}{Dodaro et~al\mbox{.}}{2018}]{DBLP:conf/aiia/DodaroGMP18}
{\sc Dodaro, C.}, {\sc Galat{\`{a}}, G.}, {\sc Maratea, M.}, {\sc and} {\sc
  Porro, I.} 2018.
\newblock Operating room scheduling via answer set programming.
\newblock In {\em AI*IA}. LNCS, vol. 11298. Springer, 445--459.

\bibitem[\protect\citeauthoryear{Erdem, Gelfond, and Leone}{Erdem
  et~al\mbox{.}}{2016}]{DBLP:journals/aim/ErdemGL16}
{\sc Erdem, E.}, {\sc Gelfond, M.}, {\sc and} {\sc Leone, N.} 2016.
\newblock Applications of answer set programming.
\newblock {\em {AI} Magazine\/}~{\em 37,\/}~3, 53--68.

\bibitem[\protect\citeauthoryear{Falkner, Friedrich, Schekotihin, Taupe, and
  Teppan}{Falkner et~al\mbox{.}}{2018}]{DBLP:journals/ki/FalknerFSTT18}
{\sc Falkner, A.~A.}, {\sc Friedrich, G.}, {\sc Schekotihin, K.}, {\sc Taupe,
  R.}, {\sc and} {\sc Teppan, E.~C.} 2018.
\newblock Industrial applications of answer set programming.
\newblock {\em Künstliche Intelligenz\/}~{\em 32,\/}~2-3, 165--176.

\bibitem[\protect\citeauthoryear{Gebser, Kaminski, Kaufmann, Ostrowski, Schaub,
  and Wanko}{Gebser et~al\mbox{.}}{2016}]{DBLP:conf/iclp/GebserKKOSW16}
{\sc Gebser, M.}, {\sc Kaminski, R.}, {\sc Kaufmann, B.}, {\sc Ostrowski, M.},
  {\sc Schaub, T.}, {\sc and} {\sc Wanko, P.} 2016.
\newblock Theory solving made easy with clingo 5.
\newblock In {\em {ICLP} (Technical Communications)}. {OASICS}, vol.~52.
  Schloss Dagstuhl - Leibniz-Zentrum fuer Informatik, 2:1--2:15.

\bibitem[\protect\citeauthoryear{Gebser, Kaufmann, and Schaub}{Gebser
  et~al\mbox{.}}{2012}]{DBLP:journals/ai/GebserKS12}
{\sc Gebser, M.}, {\sc Kaufmann, B.}, {\sc and} {\sc Schaub, T.} 2012.
\newblock Conflict-driven answer set solving: From theory to practice.
\newblock {\em Artificial Intelligence\/}~{\em 187}, 52--89.

\bibitem[\protect\citeauthoryear{Gebser, Obermeier, Schaub, Ratsch{-}Heitmann,
  and Runge}{Gebser et~al\mbox{.}}{2018}]{DBLP:journals/tplp/GebserOSR18}
{\sc Gebser, M.}, {\sc Obermeier, P.}, {\sc Schaub, T.}, {\sc
  Ratsch{-}Heitmann, M.}, {\sc and} {\sc Runge, M.} 2018.
\newblock Routing driverless transport vehicles in car assembly with answer set
  programming.
\newblock {\em Theory and Practice of Logic Programming\/}~{\em 18,\/}~3-4,
  520--534.

\bibitem[\protect\citeauthoryear{{Gurobi Optimization, LLC}}{{Gurobi
  Optimization, LLC}}{2021}]{gurobi}
{\sc {Gurobi Optimization, LLC}}. 2021.
\newblock {Gurobi Optimizer Reference Manual}.

\bibitem[\protect\citeauthoryear{Hahn-Goldberg, Carter, Beck, Trudeau, Sousa,
  and Beattie}{Hahn-Goldberg et~al\mbox{.}}{2014}]{hahn-goldberg_dynamic_2014}
{\sc Hahn-Goldberg, S.}, {\sc Carter, M.~W.}, {\sc Beck, J.~C.}, {\sc Trudeau,
  M.}, {\sc Sousa, P.}, {\sc and} {\sc Beattie, K.} 2014.
\newblock Dynamic optimization of chemotherapy outpatient scheduling with
  uncertainty.
\newblock {\em Health Care Management Science\/}~{\em 17,\/}~4, 379--392.

\bibitem[\protect\citeauthoryear{Heshmat and Eltawil}{Heshmat and
  Eltawil}{2021}]{Heshmat2021}
{\sc Heshmat, M.} {\sc and} {\sc Eltawil, A.} 2021.
\newblock Solving operational problems in outpatient chemotherapy clinics using
  mathematical programming and simulation.
\newblock {\em Annals of Operations Research\/}~{\em 298}, 1--18.

\bibitem[\protect\citeauthoryear{Huang, Bryce, Culbertson, Connor, and
  Looker}{Huang et~al\mbox{.}}{2017}]{huang_alternative_2017}
{\sc Huang, Y.-L.}, {\sc Bryce, A.~H.}, {\sc Culbertson, T.}, {\sc Connor,
  S.~L.}, {\sc and} {\sc Looker, S. A. e.~a.} 2017.
\newblock Alternative {Outpatient} {Chemotherapy} {Scheduling} {Method} to
  {Improve} {Patient} {Service} {Quality} and {Nurse} {Satisfaction}.
\newblock {\em Journal of Oncology Practice\/}~{\em 14,\/}~2, 82--91.

\bibitem[\protect\citeauthoryear{Huggins, Claudio, and Pérez}{Huggins
  et~al\mbox{.}}{2014}]{huggins_2014}
{\sc Huggins, A.}, {\sc Claudio, D.}, {\sc and} {\sc Pérez, E.} 2014.
\newblock Improving resource utilization in a cancer clinic: An optimization
  model.
\newblock In {\em {IIE Annual Conference and Expo 2014}}.

\bibitem[\protect\citeauthoryear{Ignatiev, Morgado, and
  Marques{-}Silva}{Ignatiev
  et~al\mbox{.}}{2019}]{DBLP:journals/jsat/IgnatievMM19}
{\sc Ignatiev, A.}, {\sc Morgado, A.}, {\sc and} {\sc Marques{-}Silva, J.}
  2019.
\newblock {RC2:} an efficient maxsat solver.
\newblock {\em J. Satisf. Boolean Model. Comput.\/}~{\em 11,\/}~1, 53--64.

\bibitem[\protect\citeauthoryear{Kumar and Dey}{Kumar and
  Dey}{2020}]{kumar_treatment_2020}
{\sc Kumar, D.} {\sc and} {\sc Dey, T.} 2020.
\newblock Treatment delays in oncology patients during {COVID}-19 pandemic: {A}
  perspective.
\newblock {\em Journal of Global Health\/}~{\em 10,\/}~1.
\newblock International Society of Global Health.

\bibitem[\protect\citeauthoryear{Lamé, Jouini, and Cardinal}{Lamé
  et~al\mbox{.}}{2016}]{Lame_2016}
{\sc Lamé, G.}, {\sc Jouini, O.}, {\sc and} {\sc Cardinal, J.} 2016.
\newblock Outpatient chemotherapy planning: a literature review with insights
  from a case study.
\newblock {\em IIE Transactions on Healthcare Systems Engineering\/}~{\em
  6,\/}~3, 127--139.

\bibitem[\protect\citeauthoryear{Martins, Manquinho, and Lynce}{Martins
  et~al\mbox{.}}{2014}]{DBLP:conf/sat/MartinsML14}
{\sc Martins, R.}, {\sc Manquinho, V.~M.}, {\sc and} {\sc Lynce, I.} 2014.
\newblock Open-wbo: {A} modular maxsat solver,.
\newblock In {\em {SAT} 2014}. LNCS, vol. 8561. Springer, 438--445.

\bibitem[\protect\citeauthoryear{Morgado, Dodaro, and Marques{-}Silva}{Morgado
  et~al\mbox{.}}{2014}]{DBLP:conf/cp/MorgadoDM14}
{\sc Morgado, A.}, {\sc Dodaro, C.}, {\sc and} {\sc Marques{-}Silva, J.} 2014.
\newblock {Core-Guided MaxSAT with Soft Cardinality Constraints}.
\newblock In {\em {CP} 2014}. Springer, Lyon, France, 564--573.

\bibitem[\protect\citeauthoryear{{Olivier Roussel and Vasco
  Manquinho}}{{Olivier Roussel and Vasco Manquinho}}{2012}]{pbcompetition}
{\sc {Olivier Roussel and Vasco Manquinho}}. 2012.
\newblock {Input/Output Format and Solver Requirements for the Competitions of
  Pseudo-Boolean Solvers}.

\bibitem[\protect\citeauthoryear{Sch{\"{u}}ller}{Sch{\"{u}}ller}{2018}]{DBLP:journals/ki/Schuller18}
{\sc Sch{\"{u}}ller, P.} 2018.
\newblock Answer set programming in linguistics.
\newblock {\em K{\"{u}}nstliche Intelligence\/}~{\em 32,\/}~2-3, 151--155.

\bibitem[\protect\citeauthoryear{Sevinc, Sanli, and Goker}{Sevinc
  et~al\mbox{.}}{2013}]{sevinc_algorithms_2013}
{\sc Sevinc, S.}, {\sc Sanli, U.~A.}, {\sc and} {\sc Goker, E.} 2013.
\newblock Algorithms for scheduling of chemotherapy plans.
\newblock {\em Computers in Biology and Medicine\/}~{\em 43,\/}~12, 2103--2109.

\bibitem[\protect\citeauthoryear{Sud, Jones, Broggio, Loveday, and Torr}{Sud
  et~al\mbox{.}}{2020}]{sud_collateral_2020}
{\sc Sud, A.}, {\sc Jones, M.~E.}, {\sc Broggio, J.}, {\sc Loveday, C.}, {\sc
  and} {\sc Torr, B. e.~a.} 2020.
\newblock Collateral damage: the impact on outcomes from cancer surgery of the
  {COVID}-19 pandemic.
\newblock {\em Annals of Oncology\/}~{\em 31,\/}~8, 1065--1074.
\newblock Elsevier.

\bibitem[\protect\citeauthoryear{Turkcan, Zeng, and Lawley}{Turkcan
  et~al\mbox{.}}{2012}]{turkanscheduling2010}
{\sc Turkcan, A.}, {\sc Zeng, B.}, {\sc and} {\sc Lawley, M.} 2012.
\newblock Chemotherapy operations planning and scheduling.
\newblock {\em IIE Transactions on Healthcare Systems Engineering\/}~{\em
  2,\/}~1, 31--49.

\end{thebibliography}

\clearpage

\end{document}